\documentclass[iop]{emulateapj}                                        

\usepackage{times}

\shorttitle{\textit{Hi-C} Observations of Coronal Loops}
\shortauthors{Brooks et al.}

\begin{document}


\title{High Spatial Resolution Observations of Loops in the Solar Corona}

\author{David H. Brooks\altaffilmark{1,4}, Harry P. Warren\altaffilmark{2}, Ignacio
  Ugarte-Urra\altaffilmark{1}, Amy R. Winebarger\altaffilmark{3}}

\affiliation{\altaffilmark{1}College of Science, George Mason University, 4400 University Drive,
  Fairfax, VA 22030}

\affiliation{\altaffilmark{2}Space Science Division, Naval Research Laboratory, Washington, DC
  20375}

\affiliation{\altaffilmark{3}NASA Marshall Space Flight Center, ZP 13, Huntsville, AL 35812}

\altaffiltext{4}{Current address: Hinode Team, ISAS/JAXA, 3-1-1 Yoshinodai, Chuo-ku, Sagamihara,
  Kanagawa 252-5210, Japan}


\begin{abstract}
  Understanding how the solar corona is structured is of fundamental importance to determining how
  the Sun's upper atmosphere is heated to high temperatures. Recent spectroscopic studies have
  suggested that an instrument with a spatial resolution of 200\,km or better is necessary to
  resolve coronal loops. The \textit{High Resolution Coronal Imager (Hi-C)} achieved this
  performance on a rocket flight in July 2012. We use \textit{Hi-C} data to measure the Gaussian
  widths of 91 loops observed in the solar corona and find a distribution that peaks at about
  270\,km. We also use \textit{Atmospheric Imaging Assembly (AIA)} data for a subset of these
  loops and find temperature distributions that are generally very narrow. These observations
  provide further evidence that loops in the solar corona are often structured at a scale of
  several hundred kilometers, well above the spatial scale of many proposed physical mechanisms.
\end{abstract}


 
 \section{Introduction}

 Spatial resolution plays a critical role in interpreting observations of the solar corona. At
 very low spatial resolution ($\sim 5000$\,km) it is clear that many independent structures are
 being observed along the line of sight and any attempt to model such observations must account
 for this superposition. An important question in solar physics is at what spatial scale do we
 begin to observe isothermal structures that evolve on timescales comparable to a radiative
 cooling time. Some mechanisms that are thought to be responsible for the heating of the solar
 corona are expected to operate at very small spatial scales. For example, many numerical
 simulations of magnetic reconnection \citep[e.g,][]{shay2001} suggest that current sheets form on
 the scale of the ion inertial length ($d_i = c/\omega_{pi}$), which is on the order of several
 hundred meters in the solar corona. This would suggest that the observation of resolved coronal loops
 may not be achieved for the foreseeable future.

 There is some evidence, however, that coronal loops are actually structured at much larger
 spatial scales. Some of the highest spatial resolution observations of the solar corona obtained
 on a routine basis have been from the \textit{Transition Region and Coronal Explorer}
 \citep{handy_etal1999}. Analysis of million degree loop structures observed by \textit{TRACE} in
 the \ion{Fe}{9} 171\,\AA\ channel yielded a mean width and standard deviation of 1400$\pm$300\,km
 for the distribution
 \citep{aschwanden&nightingale_2005}. This was higher than the instrumental point spread function
 of about 900\,km and suggested that loops might actually be resolved at this spatial
 resolution. Furthermore, spectroscopic observations have indicated that the temperature
 distributions in these loops are often very narrow \citep{delzanna_2003,warren_etal2008a},
 consistent with the interpretation that loops at coronal temperatures are, or are close to being,
 resolved.

 The temporal evolution of loops observed at million degree temperatures provides additional
 insights into the structuring of the corona. Observations often show loops that appear in
 progressively lower ionization stages over time, suggesting that they are cooling
 \citep[e.g.][]{winebarger_etal2003a,ugarte-urra2009,mulumoore_etal2011,viall2011}. Numerical
 simulations of this evolution, however, indicate that the loops evolve on time scales much longer
 than a radiative cooling time, suggesting that they may consistent of at least a few unresolved
 strands \citep{reale2000,warren2003}. \cite{brooks_etal2012} used densities measured from
 relatively low spatial resolution spectroscopic observations to infer the actual emitting volume
 in coronal loops. They found that the observed emission could be reproduced with only a small
 number of sub-resolution threads several hundred kilometers in width.

 Observations at optical wavelengths also suggest that coronal loops are structured on scales
 below 1000\,km but above the very small spatial scale of many proposed physical mechanisms. For
 example, the observation of coronal condensations, which form when coronal plasma cools
 catastrophically to very low temperatures and falls back to the surface of the Sun, suggests that
 coronal loops have widths of a few hundred km \citep{antolin&rouppevandervoort_2012}. The low
 temperatures of the condensations allow them to be imaged at very high spatial resolution by optical
 telescopes. The diffraction limit of the CRisp Imaging SpectroPolarimeter (CRISP) used for these coronal rain
 observations is approximately $100$\,km
 \citep{scharmer2008}. \cite{antolin&rouppevandervoort_2012} also reported condensations forming
 nearly simultaneously on adjacent field lines, consistent with the idea that heating is coherent
 across several threads over larger scales in the corona.

 \begin{figure*}
   \centerline{\includegraphics[width=0.95\linewidth,viewport=65 40 340 342]{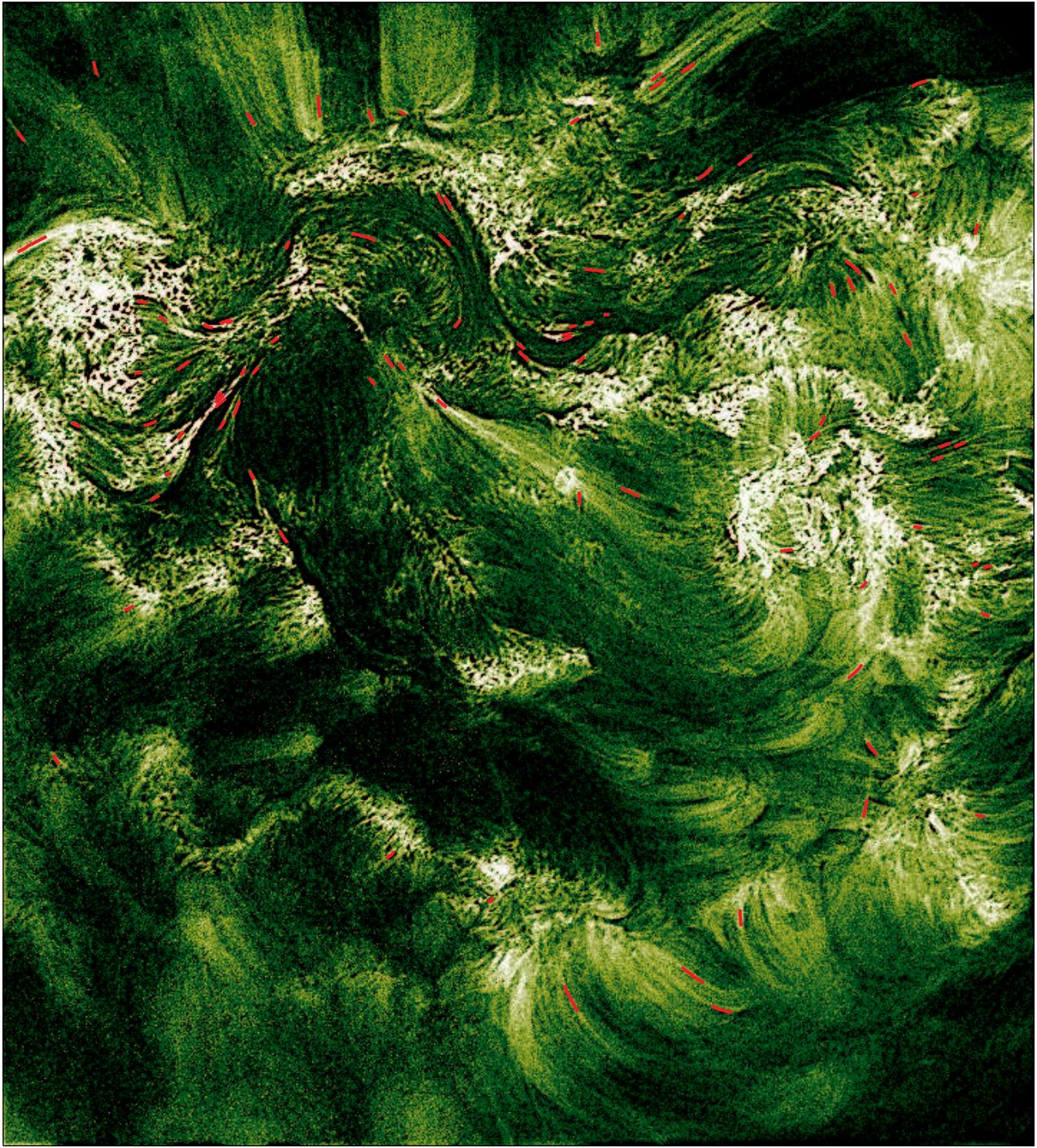}}
   \caption{\textit{Hi-C} image of AR 11520/11521 taken at 18:55:30UT on 2012, July 11. The 91
     loop segments are marked in red. The image in this figure has been treated with a Gaussian sharpening filter 
     to highlight the fine structure.}
   \label{fig1}
 \end{figure*}

 The launch of the \textit{High Resolution Coronal Imager (Hi-C)} on a sub-orbital rocket flight
 has provided a new opportunity to observe the spatial structuring of the solar corona directly
 \citep{cirtain_etal2013}. \textit{Hi-C} observed a bandpass dominated by the \ion{Fe}{12}
 195.119\,\AA\ line with a spatial resolution of about 150\,km, possibly sufficient to resolve
 loops. In this paper we examine loop cross sections for a sample of coronal loops observed with
 \textit{Hi-C} and find a distribution of Gaussian widths that peaks at about 270\,km. These
 observations provide convincing evidence that the solar corona is structured at a scale of
 several hundred kilometers, suggesting that coronal loops may be routinely resolved by the next
 generation of solar instrumentation.

 \section{Observations and Analysis Methods}

 \begin{figure*}
   \centerline{\includegraphics[width=0.95\linewidth,viewport=80 345 535 565]{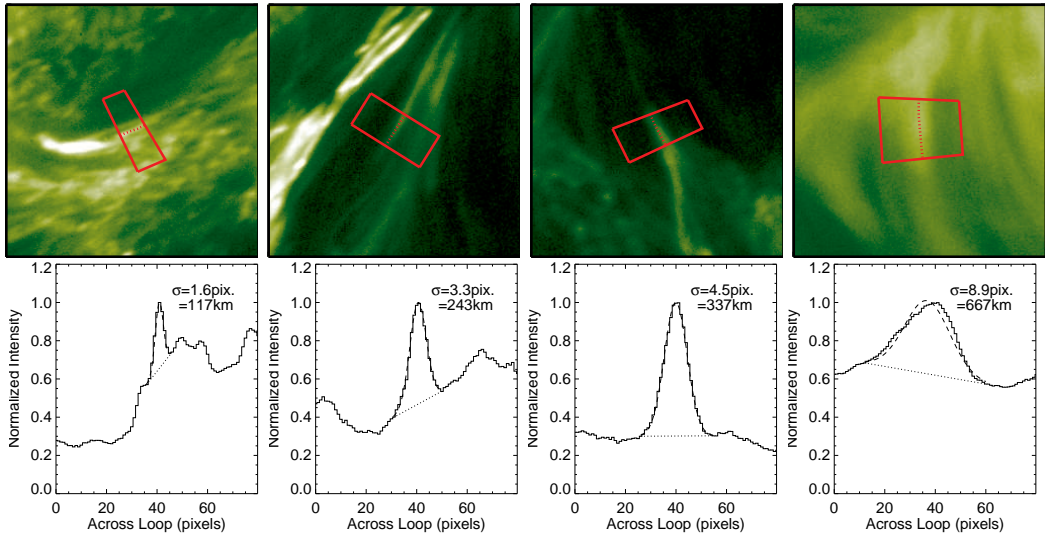}}
   \caption{Example loop segments. Top row: \textit{Hi-C} images with the loop segments
     highlighted in red. Bottom row: cross loop normalized intensity profiles (solid histogram)
     with Gaussian fits (dash) and backgrounds (dot) overlaid. The Gaussian width in \textit{Hi-C}
     pixels and $km$ is indicated in the legend. The interpolated data have been resampled to show
     the instrument pixel scale.}
 \label{fig2}
 \end{figure*}

 \begin{figure*}
 \centerline{\includegraphics[width=0.45\linewidth]{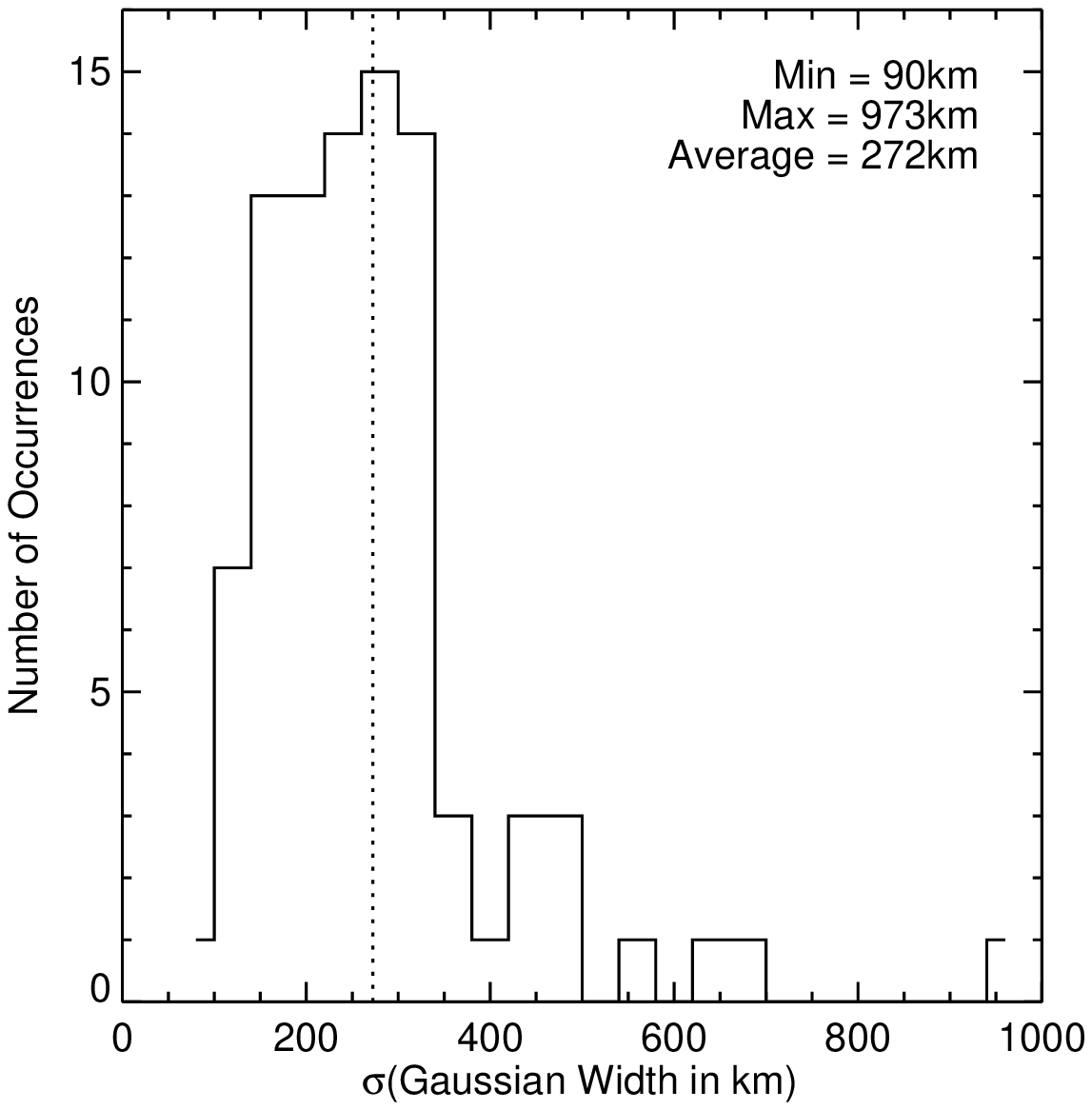}	
   \includegraphics[width=0.45\linewidth]{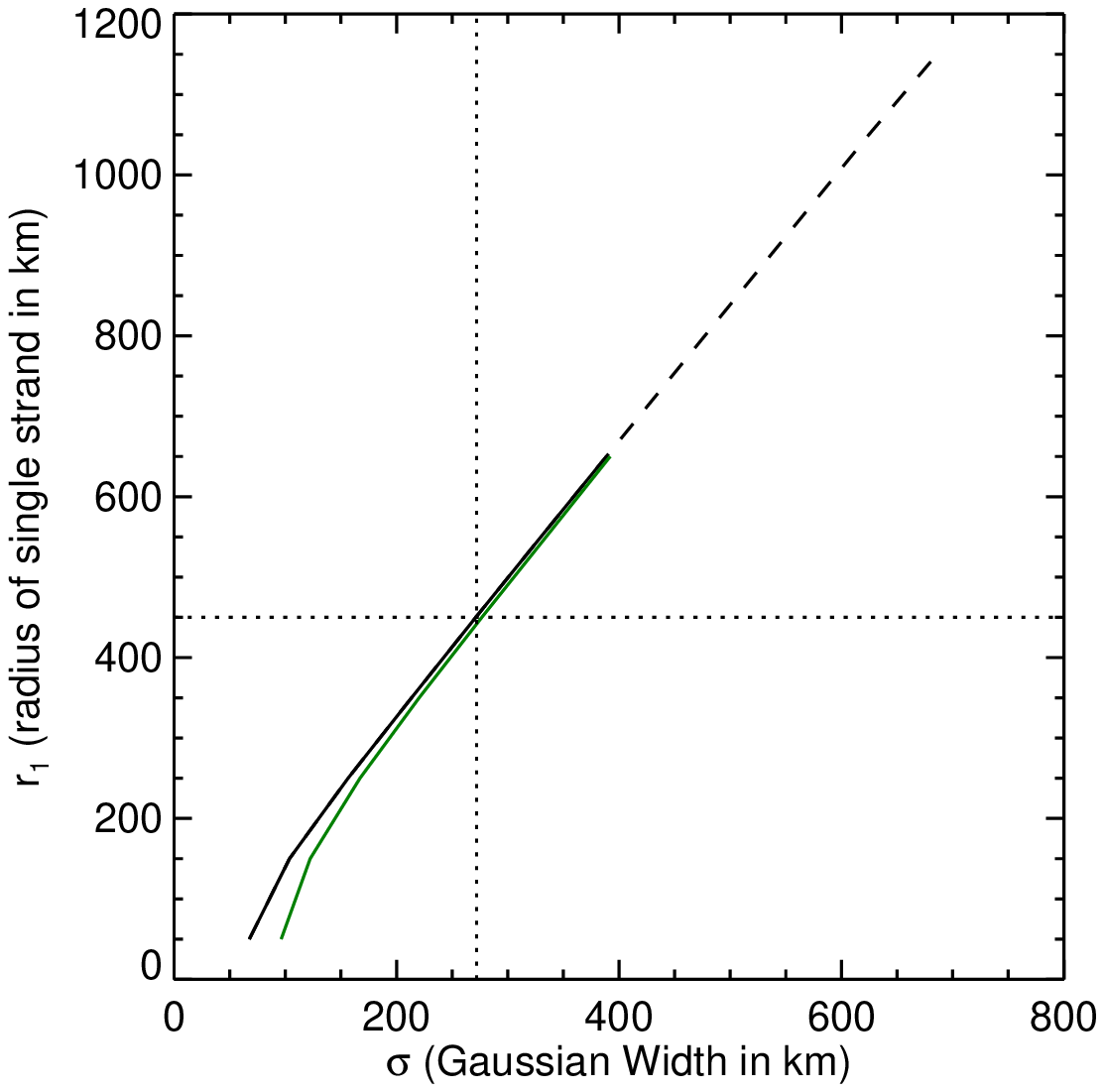}}
 \caption{Left panel: distribution of Gaussian widths measured for the sample
   of 91 loop segments shown in Figures \ref{fig1} and \ref{fig2}. Right
   panel: relationship between the true loop radius and the simulated
   \textit{Hi-C} measured Gaussian width if the loop consists of a single
   strand. As the radius becomes much larger than the PSF of the instrument,
   the Gaussian shape becomes a poorer fit to the loop cross-section (dashed
   line). Black and green lines represent a PSF of 0.2\arcsec\ and 0.3\arcsec,
   respectively.}
 \label{fig3}
 \end{figure*}

 The \textit{Hi-C} instrument comprises a CCD camera and Ritchey-Chretien telescope that observes
 in a 5\,\AA\, wide bandpass around 193\,\AA. It was flown on 2012 July 11, and obtained images of
 the solar corona at the highest spatial resolution ever achieved. Details of the instrument are
 given by \citet{cirtain_etal2013}, who indicate that the spatial resolution was close to
 0.2\arcsec\ (or about 150\,km). The target of the flight was the active region complex designated
 11520/11521. We obtained the data pre-processed from the Virtual Solar Observatory. They are full
 resolution level 1.5 data and have had the dark current subtracted. They have also been cropped,
 flat-fielded, normalized, unrolled, internally coaligned, and cleaned of dust. We also coaligned
 them to near simultaneous full Sun 193\,\AA\, images from the \textit{Atmospheric Imaging
   Assembly} \citep[\textit{AIA},][]{lemen_etal2012} on the \textit{Solar Dynamic Observatory}
 \citep[\textit{SDO},][]{pesnell_etal2012}.

 An example \textit{Hi-C} image taken at 18:55:30UT is shown in Figure~\ref{fig1}. The image has 
 has been treated with a Gaussian sharpening filter to highlight the fine structure. This filter
 was only applied for presentation, all the analysis was performed on the original data. We visually
 identified a number of loops in the 18:55:30UT image. Our sample does not cover every loop-like feature in
 the image, and there is likely to be a selection bias towards loops that are distinctive and have
 relatively pronounced cross-loop intensity profiles. Nevertheless, we were able to identify 91
 loop-like segments and they are marked in red. Following \citet{aschwanden_etal2008} and
 \citet{warren_etal2008a}, we extracted cross-loop intensity profiles by interpolating along the
 axis of the loop, straightening it, and averaging the intensities along the selected segment. We
 then selected two background positions in the intensity profile and fit them with a first order
 polynomial. A Gaussian function is then fitted to the background subtracted intensity profile and
 the Gaussian width is taken as our measurement. Several examples are shown in Figure~\ref{fig2}.

 \begin{figure}[t!]
   \centerline{\includegraphics[viewport= 60 20 535 822,clip,height=8.5in]{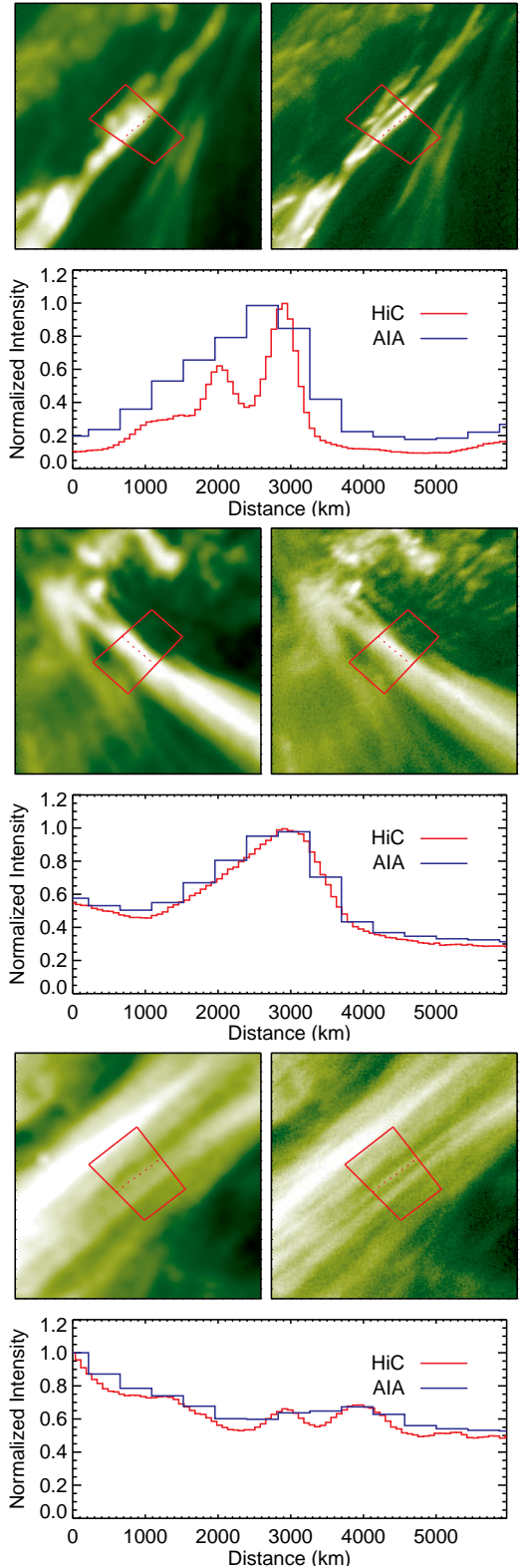}}
   \caption{\textit{Hi-C} and \textit{AIA} images and intensity profiles. Top panel: Example of a
     loop segment that is apparently composed of 1-2 structures when imaged by \textit{AIA}, but
     is revealed by \textit{Hi-C} to consist of at least 3 structures.  Middle panel: Example of a
     loop segment that appears to be a single structure when imaged by either instrument. Bottom
     panel: Example of a relatively long loop that that shows substructure in \textit{Hi-C}.}
 \label{fig4}
 \end{figure}

 Many of these loops are also visible in the \textit{AIA} images. We extracted co-spatial cross
 loop intensity profiles from the coaligned \textit{AIA} images and found that 79 of the AIA
 193\,\AA\, profiles are highly correlated (C$>$0.8) with those of \textit{Hi-C}, allowing us to
 perform an emission measure (EM) analysis of the loops using the \textit{AIA} filters. We
 therefore extracted the intensities from all the filters by fitting a Gaussian to the cross-loop
 \textit{AIA} profiles using the \textit{Hi-C} defined background positions. We then fit a
 Gaussian EM to the \textit{AIA} intensities, assuming an uncertainty of 25\%
 \citep{boerner_etal2012}.  Only the intensities of highly correlated filter profiles are used in
 the analysis, though we relax the condition a little (C$>$0.7) to allow for any residual offsets
 in the coaligned data \citep[as discussed by][]{aschwanden_etal2011}. The other intensities are
 set to zero, and we also set the 94\,\AA\, intensity to zero. This step is necessary to provide a
 high temperature constraint on the EM, which has relatively large uncertainties due to poor
 temperature fidelity and coarse coverage. We assume that the emission measure distribution is a
 Gaussian 
 \begin{equation}
   \xi(T) = \frac{EM_0}{\sigma_T\sqrt{2\pi}}
   \exp\left[-\frac{(T-T_0)^2}{2\sigma_T^2}\right],
 \end{equation}
 and use a least-squares approach to determine the best-fit total emission measure ($EM_0$), peak
 temperature ($T_0$), and thermal width ($\sigma_T$).

 \section{Results and Discussion}

 The distribution of Gaussian loop widths for the entire sample of 91 loop segments is shown in
 Figure~\ref{fig3}. The values range from 90--1000\,km. The minimum measurement of 90\,km translates
 to a 212\,km FHWM, which is about 0.12\arcsec. This is close to the laboratory performance of
 \textit{Hi-C} \citep{cirtain_etal2013}.  \citet{lopezfuentes_etal2006} and
 \citet{brooks_etal2012} have pointed out that there is a complex relationship between measured
 loop widths and the actual loop radius that depends on the instrument characteristics. This makes 
 it difficult to accurately infer the physical width from the measured width. Our view
 is that the best approach is to construct physical models of the emission, convolve them with the
 instrument parameters, and compare the results with the observations. To illustrate this we
 assume that a loop consists of a single strand at a temperature of 1.6\,MK and density of $\log
 N$ = 9 and that the plate scale is 0.07\arcsec\ and the width of the point spread function is
 0.2\arcsec\ \citep[FWHM,][]{cirtain_etal2013,kobayashi_etal2013}. As is shown in Figure~\ref{fig3},
 with these assumptions, the mean value of the distribution (272\,km) corresponds to a loop
 radius ($\sim$450\,km); a value close to the mean of the loop sample analyzed by
 \citet{brooks_etal2012}.  

 Figure~\ref{fig4} contrasts the \textit{AIA} view of three of the loop segments with the same
 region observed by \textit{Hi-C}. The loop in the top panel appears to be composed of a dominant
 bright structure and another nearby that is not spatially separated nor fully distinct in the
 cross-loop intensity profile. Not only does \textit{Hi-C} begin to separate these structures, but
 it reveals the presence of a third. 

 It is tempting to conjecture from these examples that moving to higher spatial resolution will
 always reveal new substructure, as expected on theoretical grounds. 
 This behavior, however, is very similar to that reported by
 \citet{brooks_etal2012} when comparing \textit{AIA} data with lower spatial resolution
 observations from the EUV Imaging Spectrometer \citep[\textit{EIS},][]{culhane2007} on
 \textit{Hinode}. They argued that the observation of resolved loops, and EIS structures that
 appear to separate out spatially into individual loops when imaged by \textit{AIA}, could mean
 that the spatial scales of loops are not far below the resolutions of those instruments. In turn
 suggesting a topologically simple corona or that magnetic
 braiding takes place on unexpectedly large spatial scales. The latter interpretation has been
 independently suggested by \citet{cirtain_etal2013}, who claim to have observed spatially
 resolved magnetic braids for the first time with \textit{Hi-C}. Note that the loop segment in
 Figure~\ref{fig4} is the same example as shown by \citet{cirtain_etal2013}.

 Even at the modest spatial resolution of \textit{EIS} (1800\,km FWHM PSF) some loops appear to
 be resolved \citep{brooks_etal2012}.  This indicates that some of the loops observed by
 \textit{TRACE} were probably also resolved, as suggested by
 \citet{aschwanden&nightingale_2005}. Though the latter could not be unambiguously confirmed
 because of a lack of spectroscopic diagnostics (densities).  Since \textit{Hi-C} is also an
 imager, we cannot confirm this here. We do, however, find clear cases of relatively large loops
 observed by \textit{AIA} that look almost identical at the higher spatial resolution of
 \textit{Hi-C}.  An example is shown in the middle panel of Figure~\ref{fig4}. This loop has
 a Gaussian width of 554\,km\, in \textit{AIA} and 562\,km\, when measured by \textit{Hi-C}. As
 discussed, many coronal loops observed spectroscopically at lower spatial resolution have been
 found to have narrow temperature distributions.  These loops likely correspond to examples of
 large loops observed by \textit{Hi-C} like this one. From our EM analysis we found that this loop
 has a narrow thermal width of $\sigma_T$=0.32\,MK.

 Nevertheless, the thermal width of even this loop is larger than the majority of the loops in the
 sample. The loop in the top panel of Figure~\ref{fig4}, for example, has a thermal width of
 $\sigma_T$=0.04\,MK. From our EM analysis of the complete sample of all the loops detectable with
 \textit{AIA}, we found that 70\% have $\sigma_T\lesssim$0.32\,MK. These results are comparable to
 spectroscopic measurements by EIS \citep{warren_etal2008a}, though there are relatively more
 loops with a broad temperature distribution, reflecting the larger uncertainties in the
 \textit{AIA} EM analysis. The results do, however, support the idea that these loops have narrow
 temperature distributions and are composed of only a few magnetic threads.

 Recently \citet{peter_etal2013} analyzed several long loops observed by \textit{Hi-C} and found
 that they have smooth cross-field intensity profiles. They suggested that they are either single
 monolithic structures or are composed of many very small strands. With our larger sampling of  
 the data we do in fact find some cases of long loops that show evidence of substructure when
 imaged by \textit{Hi-C}.  The loop shown in the bottom panel of Figure~\ref{fig4} is the best
 example. The results for these loops are not atypical. The widths of loops like this fall within
 the distribution of Figure~\ref{fig3}. We do agree with the conclusion of \citet{peter_etal2013}
 that many of the narrow features within the \textit{Hi-C} field of view appear to be relatively
 short loops.

 The instrument performance achieved by the \textit{Hi-C} team demonstrates that 100\,km\, spatial
 scales will be routinely observable with future coronal spectrometers such as that planned for
 \textit{Solar-C} \citep{teriaca_etal2012}.  The results presented here are encouraging to the
 view that such instruments will be able to measure the true plasma properties of coronal loops,
 and provide realistic constraints for coronal heating models. If confirmed by these instruments,
 then our \textit{Hi-C} loop width measurements already represent the true spatial scales of
 coronal structures. Theoretical models need to explain why the corona is structured on these 
 scales, and why the temperature distributions are narrow over scales of hundreds of km.


 \acknowledgments We acknowledge the High resolution Coronal Imager instrument team for making the
 flight data publicly available. MSFC/NASA led the mission and partners include the Smithsonian
 Astrophysical Observatory in Cambridge, Massachusetts; Lockheed Martin's Solar Astrophysical
 Laboratory in Palo Alto, California; the University of Central Lancashire in Lancashire, England;
 and the Lebedev Physical Institute of the Russian Academy of Sciences in Moscow. This work was
 performed under contract with the Naval Research Laboratory and was funded by the NASA {\it
   Hinode} program.  Hinode is a Japanese mission developed and launched by ISAS/JAXA, with NAOJ
 as domestic partner and NASA and STFC (UK) as international partners. It is operated by these
 agencies in co-operation with ESA and NSC (Norway).


\end{document}